\documentclass[square]{ws-procs11x85}
\usepackage{multicol,float}

\begin{document}

\title{ON THE CYLINDRICAL GRAD-SHAFRANOV EQUATION}

\author{V. S. BESKIN$^*$ and E. E. NOKHRINA}

\address{I.E.Tamm Theoretical Department, P.N.Lebedev Physical Institute,
Moscow, Russia\\
$^*$E-mail: beskin@lpi.ru\\
www.tamm.lpi.ru}

\begin{abstract}
The goal of this presentation is in paying attention to the
1D cylindrical version of the Grad-Shafranov (GS) equation. In our opinion,
this approach is more rich than classical self-similar ones, 
and more suitable for astrophysical jets we  
observe. In particular, it allows us describing the central 
(and, hence, the most energetic) part of the flow. 
  
\end{abstract}


\bodymatter

\begin{multicols}{2}
\section{Introduction}\label{aba:sec1}

An activity of many compact objects -- Active Galactic Nuclei (AGNs), Young Stellar 
Objects (YSOs), microquasars -- is associated with the highly collimated jets. These jets 
are thought to be a natural outlet of an excess angular momentum of a central 
object and accreting matter\cite{Heyvaerts-96}. The latest observations indicating 
the jet rotation in AGNs~\cite{Y07} and YSOs~\cite{Bacciotti-07} support this idea. 
The most attractive model for such outflows is the 
MHD one\cite{Heyvaerts-96, bp, pp}.

Of course, the main question within this model is the collimation 
itself~\cite{bp, pp, st, Shu-94, op}. We assume here that collimation is due 
to a finite external gas and/or magnetic pressure\cite{Appl-Camenzind-93, 
Lery-99, Beskin-Malyshkin-00}. Indeed, proposing that it is the external 
magnetic field $B_{\rm ext} \sim 10^{-6}$ G that plays the main role in 
the collimation, we obtain 
$r_{\rm jet} \sim R_{\rm in}\left(B_{\rm in}/B_{\rm ext}\right)^{1/2}$. 
Here $r$ is the distance from the rotational axis,
and the subscripts 'in' correspond to the values in 
the vicinity of the central object. The similar evaluation can be 
obtained for external pressure 
$p_{\rm ext} \sim B^2_{\rm ext}/8\pi$. E.g., for YSOs 
($B_{\rm in} \sim 10^3 {\rm G}$, $R_{\rm in} \sim R_{\odot}$) we obtain 
$r_{\rm jet} \sim 10^{15}$ cm, in agreement with observational data. Accordingly, 
for AGNs ($B_{\rm in} \sim 10^4$ G, $R_{\rm in} \sim 10^{13}$ cm) we have 
$r_{\rm jet} \sim 1$ pc. It means that the external media may indeed play 
important role in the collimation process.

The internal structure of cylindrical jets was considered both for 
non-relativistic~\cite{cl, hn2} and relativistic~\cite{Appl-Camenzind-93, 
clb, eichler, bg2, b3, bn, ld} flows. In particular, it was shown that for 
constant angular velocity of plasma $\Omega_{\rm F}$ it is impossible to obtain 
reasonable solution with total zero electric current~\cite{Appl-Camenzind-93}, 
but it can be constructed if the angular velocity vanishes at the jet boundary
and if the external pressure is not equal to zero~\cite{b3, Beskin-Malyshkin-00}.

Another result was obtained for relativistic and non-relativistic 
cylindrical flows~\cite{clb, hn2, eichler, bg1} is that   
the poloidal magnetic field $B_{\rm p}$ has a jet-like form
\begin{equation}
B_{\rm p} = \frac{B_0}{1 + r^{2}/r_{\rm core}^2},
\label{r-2}
\end{equation}
where $r_{\rm core} = v_{\rm in}\gamma_{\rm in}/\Omega$. But this relation 
corresponds to a very slow (logarithmic) growth of the magnetic flux function:
$\Psi(r) \propto \ln{r}$.
It means that if the jet core contains only a small part of the total
magnetic flux $\Psi_0$, the  jet boundary is to locate exponentially
far from the axis, magnetic field being too weak to be in equilibrium
with the external media. In what follows we'll try to resolve this 
contradiction.

Thus, we consider the following model: the flow crosses all 
the critical surfaces while the effects of the external media are
negligible. It allows us to use standard values of 
integrals of motion. As the supersonic wind expands, its pressure 
becomes comparable with the external gas and/or magnetic pressure. 
The interaction of a flow with external media results in well collimated 
jet which can be described by 1D cylindrical equations.  

\section{Basic equations}

\subsection{Relativistic flow}

For cylindrical flow one can write down electric 
and magnetic 
fields as well as the four-velocity of a plasma ${\bf u}$ 
in standard form
\begin{eqnarray}
 {\bf B} =  \frac{{\bf \nabla}\Psi \times {\bf e}_{\varphi}}{2\pi r}
  -\frac{2I}{r c}{\bf e}_{\varphi}, \quad
 {\bf E} =  -\frac{\Omega_{\rm F}}{2\pi c}{\bf \nabla}\Psi,
\label{defE} \\
 {\bf u} = \frac{\eta}{n}{\bf B} + \gamma(\Omega_{\rm F}r/c) {\bf e}_{\varphi}.
  \label{4a''}
\end{eqnarray}
Here $n$ is the concentration in the comoving reference frame, and
$\gamma^2 = {\bf u}^2 + 1$ is the Lorentz-factor. In other words, 
it is convenient to express all the values in terms of magnetic flux $\Psi$ 
and total electric current $I$, the angular velocity of plasma $\Omega_{\rm F}$ 
and the particle to magnetic flux ratio $\eta$ being constant on the magnetic 
surfaces: $\Omega_{\rm F} = \Omega_{\rm F}(\Psi)$, $\eta = \eta(\Psi)$. 
Accordingly, the trans-field GS equation can be rewritten as~\cite{b4}
\begin{eqnarray}
\frac{1}{r}\frac{{\rm d}}{{\rm d} r }
\left(\frac{A}{ r }\frac{{\rm d}\Psi}{{\rm d} r }\right)
+\frac{\Omega_{\rm F}}{c^2}\left(\frac{{\rm d}\Psi}{{\rm d} r }\right)^2
\frac{{\rm d}\Omega_{\rm F}}{{\rm d}\Psi}  \nonumber \\
+\frac{32\pi^4}{r ^2{\cal M}^2 c^4}
\frac{{\rm d}}{{\rm d}\Psi}\left(\frac{G}{A}\right) \nonumber \\
-\frac{64\pi^4\mu^2}{{\cal M}^2}\eta\frac{{\rm d}\eta}{{\rm d}\Psi}
- 16\pi^{3} nT \frac{{\rm d}s}{{\rm d}\Psi}=0.
\label{ap1}
\end{eqnarray}
Here $G = r ^2(E-\Omega_{\rm F} L)^2 +{\cal M}^{2}L^2c^2-{\cal M}^2 r ^{2} E^2$,
the Alfv\'enic factor is
$A = 1-\Omega_{\rm F}^2 r ^2/c^2-{\cal M}^2$,
\begin{equation}
{\cal M}^2 = \frac{4\pi\mu\eta^2}{n}
\end{equation}
is the Alfv\'enic Mach number, $\mu=m_{\rm p}c^2+m_{\rm p}w$ is the relativistic 
enthalpy, and the derivative ${\rm d}/{\rm d}\Psi$ acts on the integrals of 
motion only. Finally, the relativistic Bernoulli equation 
$u_{\rm p}^2 = \gamma^2 - u_{\varphi}^2 - 1$ has a form
\begin{eqnarray}
\frac{{\cal M}^4}{64\pi^4 r^2c^4}
\left(\frac{{\rm d}\Psi}{{\rm d}r}\right)^2 = 
\frac{K}{r^2A^2c^4 } - \mu^2\eta^2,
\label{ap4}
\end{eqnarray}
where
\begin{equation}
K= r^2(e')^2(A-{\cal M}^2)
+{\cal M}^4 r^{2}E^2-{\cal M}^{4}L^2c^2,
\label{ap5}
\end{equation}
and $e' = E-\Omega_{\rm F}L$.
Both equations contain relativistic integrals of motion
\begin{eqnarray}
E(\Psi) = \gamma\mu\eta c^2 + \frac{\Omega_{\rm F}I}{2\pi}, \quad
L(\Psi) = r u_{\hat\varphi}\mu\eta c + \frac{I}{2\pi},
\label{ap7}
\end{eqnarray}
which, as all other invariants, are to be determined from boundary and
critical conditions. E.g., for inner part of a flow $\Psi \ll \Psi_0$ with 
zero temperature one can choose $\Omega_{\rm F}(\Psi)= \Omega_0$, 
$\eta(\Psi)=\eta_0$, and
\begin{equation}
E(\Psi)= \mu\eta_0\gamma_{\rm in}c^2 + \frac{\Omega_0^2}{4\pi^2}\Psi, \quad
L(\Psi)= \frac{\Omega_0}{4\pi^2}\Psi.
\label{ProbStat-L1}
\end{equation}

Multiplying now equation (\ref{ap1}) on $2A{\rm d}\Psi/{\rm d} r $
and using equation (\ref{ap4}), one can find~\cite{b3}
\begin{eqnarray}
\left[\frac{(e')^2}{\mu^2\eta^2c^4}-1+\frac{\Omega_{\rm F}^2 r ^2}{c^2}
-A\frac{c_s^2}{c^2}\right]
\frac{{\rm d}{\cal M}^2}{{\rm d} r } = \nonumber \\
\frac{{\cal M}^6L^2}{A r ^3 \mu^2\eta^2c^2}
+\frac{\Omega_{\rm F}^2 r {\cal M}^2}{c^{2}}\left[2 - \frac{(e')^2}{A\mu^2\eta^2c^4}\right] \nonumber \\
+{\cal M}^2 \frac{e'}{\mu^2\eta^2c^4}\frac{{\rm d}\Psi}{{\rm d} r }\frac{{\rm d}e'}{{\rm d}\Psi} 
+{\cal M}^2\frac{ r ^2}{c^2}\Omega_{\rm F}\frac{{\rm d}\Psi}{{\rm d} r }
\frac{{\rm d}\Omega_{\rm F}}{{\rm d}\Psi} \nonumber \\
-{\cal M}^2 \left(1-\frac{\Omega_{\rm F}^2 r ^2}{c^2} + 2A\frac{c_s^2}{c^2}\right)
\frac{{\rm d}\Psi}{{\rm d} r }\frac{1}{\eta}\frac{{\rm d}\eta}{{\rm d}\Psi} \nonumber \\
-\left[\frac{A}{n}\left(\frac{\partial P}{\partial s}\right)_n
+\left(1-\frac{\Omega_{\rm F}^2 r ^2}{c^2}\right)T\right]
\frac{{\cal M}^2}{\mu}\frac{{\rm d}\Psi}{{\rm d}r}
\frac{{\rm d}s}{{\rm d}\Psi}, 
\label{ap10} 
\end{eqnarray}
where $c_s \ll c$ is the sound velocity, and the entropy $s = s(\Psi)$ is the fifth 
integral of motion. Together with Bernoulli equation (\ref{ap4}) it forms 
the system of two ordinary differential equations for Mach number ${\cal M}^2$ 
and magnetic flux $\Psi$ describing cylindrical relativistic jet.
Clear boundary conditions are
\begin{eqnarray}
\Psi(0) & = & 0, \\
P(r_{\rm jet}) & = & P_{\rm ext},
\end{eqnarray}  
where $P = B^2/8\pi + p$ is the total pressure. Determining the functions 
${\cal M}^2(r)$ and $\Psi(r)$, one can find the jet radius $r_{\rm jet}$ as well 
as the profile of the current $I$, particle energy, and toroidal component of 
the four-velocity using standard expressions 
\begin{eqnarray}
\frac{I}{2\pi} & = & \frac{L-\Omega_{\rm F}r^{2}E/c^2}
{1-\Omega_{\rm F}^{2}r^{2}/c^2-{\cal M}^{2}},
\label{p33} \\
\nonumber \\
\gamma & = & \frac{1}{\mu\eta c^2} \, \frac{(E-\Omega_{\rm F}L)-{\cal M}^{2}E}
{1-\Omega_{\rm F}^{2}r^{2}/c^2-{\cal M}^{2}},
\label{p34} \\
\nonumber \\
u_{\hat\varphi} & = & \frac{1}{\mu\eta r c} \, \frac{(E-\Omega_{\rm F}L)
 \Omega_{\rm F}r^{2}/c^2-L{\cal M}^{2}}{1-\Omega_{\rm F}^{2}r^{2}/c^2-{\cal M}^{2}}.
\label{p35}
\end{eqnarray}
 
\subsection{Nonrelativistic flow}

In the nonrelativistic limit electric and magnetic fields 
are determined by general expressions (\ref{defE}).
On the other hand, equation (\ref{4a''}) can be rewritten as  
\begin{eqnarray}
 {\bf v} & = & \frac{\eta_{\rm n}}{\rho_{\rm m}}{\bf B}+\Omega_{\rm F}
  r{\bf e}_{\varphi},
  \label{4a}
\end{eqnarray}
where $\rho_{\rm m} = m_{\rm p}n$ is the mass density and 
$\eta_{\rm n}(\Psi)$ is nonrelativistic particle to 
magnetic flux ratio. Accordingly, nonrelativistic fluxes of energy 
$E_{\rm n}$ and $z$ component of the angular momentum $L_{\rm n}$
are 
\begin{eqnarray}
  E_{\rm n}(\Psi) & = & \frac{\Omega_{\rm F} I}{2\pi c\eta_{\rm n}}
  +\frac{v^2}{2} + w,
  \label{5a}  \\
   L_{\rm n}(\Psi) & = & \frac{I}{2\pi c\eta_{\rm n}}+v_{\varphi}r\sin\theta.
    \label{6a}
\end{eqnarray}   
Further, algebraic relations (\ref{p33})--(\ref{p35}) can be rewritten 
as
\begin{eqnarray}
 \frac{I}{2\pi} & = & c\eta_{\rm n}
\frac{L_{\rm n}-\Omega_{\rm F}r^2}{1-{\cal M}^2},
\label{Inrel} \\
   v_{\varphi} & = & \frac{1}{r}\frac{\Omega_{\rm F}r^2
    -L_{\rm n}{\cal M}^2}{1-{\cal M}^2},
    \label{9a}
     \end{eqnarray}
where now
\begin{equation}
 {\cal M}^{2}=\frac{4\pi\eta_{\rm n}^{2}}{\rho_{\rm m}}.
  \label{10a}
   \end{equation}
As a result, nonrelativistic Bernoulli equation
\begin{eqnarray}
  \frac{{\cal M}^4}{64\pi^4\eta_{\rm n}^2}\left(\frac{{\rm d}\Psi}{{\rm d}r}\right)^2
    = 2r^{2} (E_{\rm n} - w) \nonumber \\
  -\frac{(\Omega_{\rm F}r^2- L_{\rm n}{\cal M}^2)^2}
{(1-{\cal M}^2)^2}
      -2r^2\Omega_{\rm F}\frac{L_{\rm n}-\Omega_{\rm F}r^2}
{1-{\cal M}^2},
       \label{11a}
        \end{eqnarray}   
together with nonrelativistic limit of equation (\ref{ap10})
\begin{eqnarray}
\displaystyle\left[\vphantom{\frac{1}{2}}2e_{\rm n}-
2w+\Omega_{\rm F}^2r^2-(1-{\cal M}^2)c_{\rm s}^2\right] \frac{{\rm d}{\cal M}^2}{{\rm d}r} = \nonumber \\
\displaystyle \frac{{\cal M}^6}{1-{\cal M}^2}\frac{L_{\rm n}^2}{r^3} 
- \frac{\Omega_{\rm F}^2r}{1-{\cal M}^2}{\cal M}^2(2{\cal M}^2-1)  \nonumber \\
\displaystyle  + {\cal M}^2\frac{{\rm d}\Psi}{{\rm d}r}\frac{{\rm d}e_{\rm n}}{{\rm d}\Psi} 
+{\cal M}^2r^2\Omega_{\rm F}\frac{{\rm d}\Psi}{{\rm d}r} \frac{{\rm d}\Omega_{\rm F}}{{\rm d}\Psi} \nonumber \\
\displaystyle  +2 \left[e_{\rm n}-w+\frac{\Omega_{\rm F}^2r^2}{2}-(1-{\cal M}^2)c_{\rm s}^2\right]
\frac{{\cal M}^2}{\eta_{\rm n}}\frac{{\rm d}\Psi}{{\rm d}r}\,\frac{{\rm d}\eta_{\rm n}}{{\rm d}\Psi} \nonumber \\
\displaystyle -{\cal M}^2\left[(1-{\cal M}^2)\frac{1}{\rho_{\rm m}}
\left(\frac{\partial P}{\partial s}\right)_{\rho_{\rm m}}
+\frac{T}{m_{\rm p}}\right]\frac{{\rm d}\Psi}{{\rm d}r} \frac{{\rm {\rm d}}s}{{\rm d}\Psi}, 
\label{GS-Me}
\end{eqnarray}
where $e_{\rm n} = E_{\rm n}-\Omega_{\rm F}L_{\rm n}$, determine the
structure of nonrelativistic cylindrical flow.

\section{Advantages}

Certainly, the approach under consideration is 1D as well. For this reason, 
it has some properties similar to another self--similar ones. In 
particular, one can easily check that the singularity on the fast magnetisonic 
surface is absent. On the other hand, singularity appears on the cusp surface 
where the factors in front of ${\rm d}{\cal M}^2/{\rm d} r$ in (\ref{ap10}) 
and (\ref{GS-Me}) vanish. Nevertheless, in our opinion, this one-dimensional 
approach has clear advantages in comparison with the standard self-similar 
ones~\cite{pp, bp, st, Shu-94}. 

First of all, it allows us to use any form of the five integrals of motion.
Indeed, the self-similarity of a flow demands definite dependence of invariants 
which may be not correspond to the real boundary conditions. E.g., for relativistic
self-similar flow the angular velocity $\Omega_{\rm F}$ is to have
the form $\Omega_{\rm F}\propto r^{-1}$~\cite{lcb}. It does not
correspond neither to the homogeneous stellar rotation,
nor to the Keplerian disk rotation. Moreover, this dependence has the
singularity at the rotational axis. Thus, the standard self-similar approach
cannot describe the central (and, hence, the most energetic) part of the 
flow. 

Further, classical self-similar approach cannot describe the region of electric 
current closure. Finally, for relativistic magnetically dominated flow 
it is more convenient to use first-order equation (\ref{ap10}) instead 
of second order GS equation for which it is necessary to be careful
in taking into account small but important terms $\sim \gamma^{-2}$.
Indeed, the force balance equation (\ref{ap10}) does not contain 
the leading terms $\rho_{\rm e}{\bf E}$ and ${\bf j} \times {\bf B}/c$ as they
are analytically removed using Bernoulli equation.
As a result, as
\begin{equation}
\frac{|\rho_{\rm e}{\bf E} + {\bf j} \times {\bf B}/c|}
{|{\bf j} \times {\bf B}/c|} \sim \frac{1}{\gamma^2},
\label{forcebls}
\end{equation}
all the terms in equation (\ref{ap10}) are of the same order.

In particular, in the limit $r \gg r_{\rm core}$, ${\cal M}^2 \gg 1$ equation 
(\ref{ap10}) can be rewritten in the simple form~\cite{Beskin-Malyshkin-00} 
\begin{equation}
\frac{{\rm d}}{{\rm d}r}
\left(\frac{\mu\eta\Omega_{\rm F}r^2}{{\cal M}^2}\right)
- \frac{{\cal M}^2L^2}
{\mu\eta \Omega_{\rm F} r^3(\Omega_{\rm F}^2r^2/c^2+{\cal M}^2)} = 0.
\label{qq}
\end{equation}
Without the last term $\propto L^2(\Psi)$ equation (\ref{qq}) results in the 
conservation of the value $H$
\begin{equation}
H = \frac{\Omega_{\rm F}\eta r^2}{{\cal M}^2} =  {\rm const}
\label{qq1}
\end{equation}
was found in~\cite{Heyvaerts-Norman-89} for conical magnetic field. It is the 
conservation of $H$ that results in the jet-like solution (\ref{r-2}). Indeed, 
as $\eta(\Psi) \approx $ const and $\Omega_{\rm F}(\Psi) \approx$ const in the 
center of a jet, we obtain ${\cal M}^2 \propto r^2$. Using now the definitions 
${\cal M}^2 = 4\pi \eta^2\mu /n$ and $n u_{\rm p} = \eta B_{\rm p}$ (and the 
condition $u_{\rm p} \approx$ const fulfilled in the very center of a flow), 
we return to (\ref{r-2}). But, as we will see, the term containing $L^2$ 
(which appears to be missed previously) can be important~\cite{b3}. It is this 
term that can change the jet-like structure in relativistic case.

\section{Internal structure of cylindrical jets}

\subsection{Relativistic flow}

\subsubsection{General properties}

The solution of equations (\ref{ap4}) and (\ref{ap10}) depends 
essentially on the Mach number on the rotational axis 
${\cal M}_0^2 = {\cal M}^2(0)$~\cite{Beskin-Malyshkin-00}. 
For ${\cal M}_0^2 \gg {\cal M}_{\rm cr}^2$ where
${\cal M}_{\rm cr}^2 = \gamma_{\rm in}^2$  
\begin{equation}
{\cal M}^{2} = 
{\cal M}_{0}^{2}\left(1 + \frac{r^2}{\gamma_{\rm in}^2 R_{\rm L}^2}\right),
\label{sol0}
\end{equation}
the poloidal magnetic field corresponding to jet-like solution (\ref{r-2}). 
On the other hand, for ${\cal M}_0^2 \ll {\cal M}_{\rm cr}^2$ 
\begin{equation}
{\cal M}^{2} = {\cal M}_{0}^{2}
\left(1 + \frac{r}{\gamma_{\rm in} R_{\rm L}}\right), \;
\Psi = \frac{\gamma_{\rm in} \Psi_0}{2{\cal M}_{0}^{2}\sigma}  
\left(\frac{r}{R_{\rm L}}\right)^2.
\label{sol1}
\end{equation}
Here $R_{\rm L} = c/\Omega_{\rm F}(0)$, and 
$\sigma = \Omega_{0}^2\Psi_{0}/8\pi^2 c^2 \mu\eta_{0}$ is the Michel 
magnetization parameter~\cite{michel69} ($\gamma \approx \sigma$ for
particle dominated flow $W_{\rm part} \approx W_{\rm em}$). 
It means that $B_{\rm p} \approx$ const, i.e., the solution has no 
jet-like form.

As was already stressed, the solution (\ref{sol0}) cannot be realized
in the presence of the external media. Hence, for any finite  
pressure $P_{\rm ext}$ magnetic field in the center of cylindrical jet 
$B_0 = 4 \pi \eta_0 \mu \gamma_{\rm in}/{\cal M}_{0}^2$
cannot be much smaller than 
$B_{\rm min} = 4 \pi \eta_0 \mu \gamma_{\rm in}/{\cal M}_{\rm cr}^2$.
It gives
\begin{equation}
B_{\rm min} = \frac{1}{\sigma\gamma_{\rm in}}B(R_{\rm L}),
\label{38}
\end{equation}
where $B(R_{\rm L}) = \Psi_0/\pi R_{\rm L}^2$. 

\subsubsection{Central core}

Thus, for external magnetic field $B_{\rm ext} > B_{\rm min}$ the internal 
structure of a relativistic jet is to be described by relations (\ref{sol1}). 
On the other hand, for $B_{\rm ext} < B_{\rm min}$ the core with 
$B_{\rm p} \approx  B_{\rm min}$ is to be formed in the center of a 
flow (i.e., for $r < \gamma_{\rm in} R_{\rm L}$). In particular, for 
$\sigma^{-2}B(R_{\rm L}) < B_{\rm ext} < B_{\rm min}$ 
(and for $r \gg \gamma_{\rm in} R_{\rm L}$) 
the solution can be presented as~\cite{bn} 
\begin{equation}
{\cal M}^{2}\propto r^{\alpha}, \quad \Psi \propto r^{\beta},
\label{a+b}
\end{equation}
the sum being $\alpha + \beta = 3$. E.g., for $B_{\rm ext} = B_{\rm min}$ 
we have $\alpha = 1$, $\beta = 2$ (cf. \ref{sol1}), and for 
$B_{\rm ext} = \sigma^{-2}B(R_{\rm L})$ we have $\alpha = 2$, $\beta = 1$.

The results presented above were reproduced recently both analytically and
numerically. In~\cite{bn} it was shown that 1D approximation 
becomes true for paraboloidal outflow at large distances from the 
equatorial plane $z \gg \sigma^{2/3}R_{\rm L}$ where the flow becomes
actually cylindrical. Up to the distance $z=\sigma\gamma_{\rm in}R_{\rm L}$ 
one can use the relations (\ref{sol1}), so that the poloidal magnetic field
does not depend on $r$. The region $z > \sigma \gamma_{\rm in} R_{\rm L}$  
corresponds to core-like solution (\ref{a+b}). Nevertheless, the transverse 
dimension of a jet remains parabolic: $r_{\rm jet} \propto z^{1/2}$. Numerically 
the scalings (\ref{a+b}) were confirmed in~\cite{kbvk}. 

Remember that the existence of cylindrical core with 
$r_{\rm core} \sim \gamma_{\rm in }R_{\rm L}$ was predicted 
in many papers~\cite{Heyvaerts-Norman-89, bg2}, but magnetic flux 
$\Psi_{\rm core} = \pi r_{\rm core}^2B_{\rm min}$ inside the core was 
unknown up to now. As we see, in relativistic case the central core contains 
only a small part of the magnetic flux:
\begin{equation}
\frac{\Psi_{\rm core}}{\Psi_0} \approx \frac{\gamma_{\rm in}}{\sigma}.
\label{39}
\end{equation}
Nevertheless, as $\beta > 0$, such core-like flow can exist in the presence 
of external media.

\subsubsection{Bulk acceleration}

As on the fast magnetosonic surface 
the bulk plasma Lorentz-factor $\gamma(r_{\rm F}) \approx \sigma^{1/3}$ (and, hence, 
here $W_{\rm part}/W_{\rm em} \sim \sigma^{-2/3} \ll 1$)~\cite{michel69, 
bkr}, the additional particle acceleration is possible for $r > r_{\rm F}$. Using 
equation (\ref{p34}) and relation $\alpha + \beta = 3$ one can find that 
in all region  $B_{\rm ext} > \sigma^{-2}B(R_{\rm L})$ ($z < \sigma^2 R_{\rm L}$ 
for parabolic flow) the Lorentz-factor for $r > \gamma_{\rm in}R_{\rm L}$ can be 
determined as 
\begin{equation}
\gamma \approx r/R_{\rm L}.
\label{gx}
\end{equation} 
Accordingly, one can write down~\cite{b3}
\begin{equation}
\frac{W_{\rm part}}{W_{\rm em}} \sim
\frac{1}{\sigma}\left[\frac{B(R_{\rm L})}{B_{\rm ext}}\right]^{1/2}.
\label{46b}
\end{equation}
It means that for  $B_{\rm ext} \sim \sigma^{-2}B(R_{\rm L})$
($z \sim \sigma^2 R_{\rm L}$ for parabolic flow) where the transverse
jet dimension $r_{\rm jet} \sim \sigma R_{\rm L}$ almost the full energy 
transformation from the Poynting to particle energy flux can be 
realized. In particular, for the particle moving along parabolic 
magnetic field line one can obtain
\begin{equation}
\gamma(z) \propto (z/R_{\rm L})^{1/2}. 
\label{gz}
\end{equation}
This scaling was confirmed numerically as well~\cite{McK, Narayan}.

It is necessary to stress that relation (\ref{gx}) takes place only if one 
can neglect the curvature of magnetic surfaces. Indeed, for magnetically 
dominated case in the limit $r \gg r_{\rm F}$ the leading terms in 2D GS 
equation can be rewritten in the simple form~\cite{bn}
\begin{equation}
-\frac{1}{2} \, {\bf n} \cdot \nabla (B_{\rm p}^2) 
- \frac{{B_{\varphi}^2}}{R_{\rm c}}
+ \frac{B_{\varphi}^2 - {\bf E}^2}{r} \, ({\bf n} \cdot {\bf e}_{r}) = 0. 
\label{balance}
\end{equation}
Here $R_{\rm c}$ is the (poloidal) curvature radius of magnetic surfaces, 
and ${\bf n} = \nabla \Psi/|\nabla \Psi|$. Neglecting now the curvature term 
and using standard relations $B_{\varphi} \approx B_{\rm p} r/R_{\rm L}$ 
and $B_{\varphi}^2 - {\bf E}^2 \approx B_{\varphi}^2/\gamma_{\rm in}^2$
resulting from (\ref{defE}) and (\ref{ap4}), we return to (\ref{gx}). On 
the other hand, if the curvature is important, then one can neglect the 
first term in (\ref{balance}), and we obtain
\begin{equation}
\gamma \approx \left(R_{\rm c}/r\right)^{1/2}.
\label{gx1}
\end{equation} 
This scaling taking place for split-monopole geometry outside the fast magnetosonic 
surface corresponds to $\gamma \approx \sigma^{1/3}\ln^{1/3}(r/r_{\rm F})$~\cite{tomi, bkr}. 
Remember that for $r < r_{\rm F}$ we have "linear" acceleration (\ref{gx}). Thus, the 
effective particle acceleration can take place only if $r_{\rm jet} \sim \sigma R_{\rm L}$,
and if the curvature of magnetic surfaces is not important. 

\subsubsection{In the center of the self-similar domain}

The approach under consideration allows us matching the self-similar solution 
to the rotational axis. Indeed, for relativistic self-similar 
invariants
\begin{eqnarray}
\Omega_{\rm F}(\Psi) & = &  \Omega_0 (\Psi/\Psi_{\rm b})^{-b},
\label{int1''} \\
E(\Psi) & = & E_0 (\Psi/\Psi_{\rm b})^{1 - 2b},
\label{int2''} \\
L(\Psi) & = & L_0 (\Psi/\Psi_{\rm b})^{1 - b},
\label{int3''} \\
\eta(\Psi) & = & \eta_0 (\Psi/\Psi_{\rm b})^{1 - 2b},
\label{int4''}
\end{eqnarray}
the solution of the 2D GS equation for $\Psi > \Psi_{\rm b}$ has the form 
$\Psi(\rho,\theta) = \rho^{1/b} \Theta(\theta)$, where $\rho$ is the spherical 
radius. Hence, far from the equatorial plane where $z \gg r$  ($\theta \ll 1$,
$\rho \approx z$) one can write down
\begin{equation}
\Psi(\rho,\theta) = {\cal A} \rho^{1/b} \theta^a.
\label{autoan}
\end{equation} 
As a result, the cylindrical radius of the boundary $\Psi = \Psi_{\rm b}$ 
can be written as
\begin{equation}
r_{\rm b}(z) = {\cal A}^{-1/a} \Psi_{\rm b}^{1/a} z^{1 - 1/ab}.
\end{equation}

Let us consider now the central part of a flow $\Psi < \Psi_{\rm b}$. If again 
$\theta \ll 1$, one can integrate 1D cylindrical equations (\ref{ap4}) and 
(\ref{ap10}) considering $z \approx \rho$ as a parameter. Assuming that 
$\Omega_{\rm F}  =\Omega_0$ and $\eta =\eta_0$ for $\Psi < \Psi_{\rm b}$ 
and using solution (\ref{sol1}) 
we have for ${\cal M}^2_{\rm b}(z) = {\cal M}^2(r_{\rm b})$ 
\begin{equation}
{\cal M}^2_{\rm b}(z) 
= \frac{8\pi^2 \eta_0 \mu}{a R_{\rm L}{\cal A}^{3/a} \Psi_{\rm b}^{1-3/a}} z^{3 -3/ab}.
\end{equation}
On the other hand, for $\Psi > \Psi_{\rm b}$ ($r > r_{\rm b}$)
one can seek the solution in a form 
${\cal M}^2(r) = {\cal M_{\rm b}}^2(r/r_{\rm b})^{\varepsilon}$.
As a result, equations (\ref{ap4}), (\ref{ap10}) give
\begin{equation}
a = 2, \quad \varepsilon = 3 - 6b.
\end{equation}
Substituting now $r \approx z \theta$, we obtain
\begin{equation}
{\cal M}^2 = {\cal C} \theta^{\varepsilon}.
\end{equation}
The coefficient 
${\cal C} \propto {\cal M_{\rm b}}^2(z) z^{\varepsilon}/r_{\rm b}^{\varepsilon}(z)$, 
in agreement with self-similar property, does not depend on $z$.

\subsection{Nonrelativistic flow}

\subsubsection{Central core}

For nonrelativistic case in the central part of a flow one can use general 
expressions (\ref{ProbStat-L1}) 
\begin{equation}
E_{\rm n}(\Psi)= \frac{v_{\rm in}^2}{2} + i_0\frac{\Omega_0^2}{4\pi^2 c\eta_0}\Psi, \quad
L_{\rm n}(\Psi)=i_0\frac{\Omega_0}{4\pi^2 c \eta_0}\Psi,
\label{ProbStat-L2}
\end{equation}
non-dimensional current $i_0 = j/j_{\rm GJ}$ depending now on the angular velocity  
$\Omega_{\rm F}$. For $\Omega_{\rm F} \ll \Omega_{\rm cr}$, where
\begin{equation}
\Omega_{\rm cr} = \frac{v_{\rm in}}{R_{\rm in}}
\left( \frac{\rho_{\rm in}v_{\rm in}^2}{B_{\rm in}^2/8\pi}\right)^{1/2},
\end{equation}
corresponding to particle dominated outflow near the star 
the 2D problem can be solved analytically~\cite{bo, bg1}, and we obtain 
$i_0 = c/v_{\rm in}$. For magnetically dominated flow  near the origin 
one can write down~\cite{Lery-99}
\begin{equation}
i_0 \approx c/v_{\rm in} \left(\Omega_{\rm F}/\Omega_{\rm cr}\right)^{-2/3}.
\end{equation}
Nevertheless, for $\Psi < \Psi_{\rm in}$, where
\begin{equation}
\Psi_{\rm in} = \frac{4 \pi^2 v_{\rm in}^3\eta_0}{i_0\Omega_0^2},
\label{inin}
\end{equation}
the flow remains particle dominated: $E_{\rm n} \approx v_{\rm in}^2/2$.
Remember that in the nonrelativistic case the flow can pass smoothly 
the critical surfaces only if $W_{\rm part}(r_{\rm F}) \sim W_{\rm em}(r_{\rm F})$.
Thus, the flow at large distances is to be particle dominated.

Solving now equations (\ref{11a}) and (\ref{GS-Me}) for sub Alfv\'enic flow 
${\cal M}_0^2 < 1$  we obtain that poloidal magnetic field remains constant 
inside the jet up to the very boundary. Thus, one can put $B(0) = B_{\rm ext}$.
But such a flow can exist only in the presence of large 
enough external magnetic field $B_{\rm ext} > B(r_{\rm F})$. For ordinary 
YSOs $B(r_{\rm F}) \sim 10^{-1}$ G, so sub Alfv\'enic flow in a jet cannot 
be realized. 

On the other hand, for super-Alfv\'enic cold outflow one can find that
the term $\propto L_{\rm n}^2$ in (\ref{GS-Me}) plays no role.
It means that here $H \approx $ const, 
and we return to jet-like solution (\ref{r-2})\cite{eichler, bg1}.
But, as was already stressed, in the presence of finite external pressure
it is possible if the central core $r < r_{\rm core} = v_{\rm in}/\Omega$
contains almost all magnetic flux $\Psi_0$. This can be realized only for slow 
rotation $\Omega_{\rm F} \ll \Omega_{\rm cr}$. In this case magnetic field
on the axis cannot be much smaller than $B_{\rm min} = \Psi_0/\pi r_{\rm core}^2$:
\begin{equation}
B_0 = \frac{B_{\rm min}}{\ln(1 + B_{\rm min}/B_{\rm ext})}.
\label{B0min}
\end{equation}
Accordingly, $\Psi_{\rm core} = \Psi_0/\ln(1 + B_{\rm min}/B_{\rm ext})$.
This structure was reproduced numerically as well~\cite{Lery-99}.

But for fast rotation $\Omega_{\rm F} \gg \Omega_{\rm cr}$
the core magnetic flux  $\Psi_{\rm core}$ is much smaller even than the 
flux $\Psi_{\rm in}$ (\ref{inin}) within the central part of a flow:
\begin{equation}
\frac{\Psi_{\rm core}}{\Psi_{\rm in}} 
\approx \frac{i_0 v_{\rm in}}{2c  M_{0}^{2}} \ll 1.
\end{equation}
It means that the cold cylindrical flow resulting from the interaction of
fast rotating supersonic wind with the external media cannot be realized.

\subsubsection{Heating in oblique shock}

To resolve this contradiction, one can propose that in real nonrelativistic jets 
an important role may play the finite temperature. E.g., the additional heating can 
be connected with the oblique shock near the base of a jet~\cite{bt, Levinson-07}. 
It is well known that such a shock is needed to explain the emission lines observed 
in YSOs~\cite{Schwartz-83}.This situation is alike the  hydrodynamical supersonic 
flow meeting the wall. This analogy is all the more reasonable as the 
non-relativistic supersonic outflow is to be particle dominated.  

To evaluate the thermal terms in equations (\ref{11a}) and (\ref{GS-Me}) we
consider pure hydrodynamical shock wave swifting spherically symmetric supersonic
flow into cylindrical jet. Knowing the swifting angle, one can determine the entropy
jump $\Delta s$ as a function of particle flux, all other four invariants being the 
same as in front of a shock. As a result, we found that fastly rotating jet with
$\Omega_{\rm F} \gg \Omega_{\rm cr}$ heated in a shock is to have core-jet structure
(\ref{a+b}) with $\alpha < 2$, $\beta > 0$. Hence, it can be realized in the presence 
of external media. Obtained jet parameters ($T \sim 10^4 \, {\rm K}$, $v_{\varphi} \sim 10$ km/s
at $r \sim 10$ A.U.)~\cite{Grenoble} are in agreement with observational data.

\subsubsection{In the center of the self-similar domain}

The procedure similar to Sect. 4.1.4. for nonrelativistic particle 
dominated flow gives for self-similar region ($\Psi > \Psi_{\rm b}$, 
$\theta \ll 1$, $E_{\rm n} \propto \Psi^{-b'}$) that 
${\cal M}^2 = {\cal C}\theta^{\varepsilon}$, $\varepsilon = 2 - 4b'$,  
as in numerical simulation~\cite{vla}.

\section{Conclusion}

Thus, cylindrical GS equation has definite advantages in comparison with 
standard self-similar ones. Using this approach it was demonstrated that 
in relativistic case effective particle acceleration can take place only 
if $r_{\rm jet} \sim \sigma R_{\rm L}$, the curvature of magnetic surfaces
playing no role. For nonrelativistic flow we found that the heating in 
oblique shock near the base of a jet must play the leading role for 
magnetically dominated flow. In both cases the magnetic flux within the 
central core was determined.

\section{Acknowledgments}

This work was supported by Russian Foundation for Basic Research 
(Grant no.~08-02-00749) and Dinasty fund.

\end{multicols}
\end{document}